\documentclass[prl,twocolumn,showpacs,preprintnumbers,amsmath,amssymb]{revtex4}

\usepackage{graphicx}
\usepackage{dcolumn}
\usepackage{bm}
\usepackage{subfigure}

\usepackage[hypertex]{hyperref}
\hypersetup{pdftitle={Detection of phase synchronization, Axel Rossberg},%
  breaklinks,%
  bookmarksopen=false,%
  pdfpagemode=None,%
}
\usepackage{amsmath}
\usepackage{amssymb}
\usepackage{graphicx}
\newcommand{\var}{\mathop{\mathrm{var}}}
\newcommand{\cov}{\mathop{\mathrm{cov}}}

\begin{document}


\title{Phase synchronization from noisy univariate
  signals}

\author{A. G. Rossberg}
\author{K. Bartholom\'{e}}
\author{H. U. Voss}
\author{J. Timmer}
\affiliation{Center for Data Analysis and Modeling,
 Albert-Ludwigs-Universit\"{a}t Freiburg, Eckerstr. 1, 79104
 Freiburg, Germany}%
\date{Phys. Rev. Lett. 93, 154103 (2004)}


\begin{abstract}
  We present methods for detecting phase synchronization of two
  unidirectionally coupled, self-sustained noisy oscillators from a signal of the
  driven oscillator alone.  One method detects soft, another hard phase
  locking. Both are applied to the problem of detecting phase
  synchronization in von K{\'a}rm{\'a}n vortex flow meters. 
\end{abstract}

\maketitle

Phase synchronization of nonlinear, self-sustained oscillators is
known since the observations of Huygens over 300 years ago
\cite{huygens1673}.  For several reasons, interest in the
phenomenon has recently resurged, for review, see
\cite{boccaletti2002,pikovsky2001,mosekilde2002}.  For
chaotic oscillators, phase synchronization was found to be an
independent regime among various others with different degrees of
synchronization
\cite{rosenblum1996,%
rosenblum1997,%
josic2001,%
drepper2000%
}.  Phase synchronization has also been observed in several natural
phenomena
\cite{boccaletti2002}%
, most notably
perhaps in brain physiology \cite{tass1998,%
bhattacharya2001,%
varela2001%
}, where it is thought to play a
key role in information processing \cite{varela2001}.
Therefore, interest has 
risen in the problem of identifying and characterizing phase
synchronization from measured time series
\cite{rosenblum2001a}.

For the situation that \emph{signals from both oscillators} are
available, several techniques have been developed to quantify the
degree of phase synchronization between them
\cite{tass1998,%
  allefeld2003}, to determine the
coupling direction \cite{rosenblum2001b,%
  rosenblum2002} and 
the type of coupling \cite{jamsek2003}.

For \emph{univariate signals} from bivariate systems it is sometimes
possible to identify and separate oscillatory components of the two
oscillators, and to proceed 
as in the bivariate case \cite{
stefanovska2000,%
jamsek2003,%
rossberg2004%
}.

But when a univariate signal contains oscillations from a single
oscillator only, the question whether it is phase-synchronized to
some other
oscillator is much harder to answer. Important progress towards
this goal is the method of Janson \emph{et.\ al.}\ \cite{%
janson2001%
}, where the
driving oscillator is identified by
observing the regular variations it induces in the 
period of the driven oscillator.  Effectively, these observations are
used to reconstruct the dynamics of the pair of coupled oscillators in
the spirit of Takens' delay embedding.  As for all embedding
techniques, the method is sensitive to noise, and careful
pre-processing of the data is required for noise reduction.

In this Letter, a different method for detecting synchronization in
terms of phase locking from univariate time series, based on 
nonequilibrium phase statistics, is proposed.  It applies to
situations with unidirectional coupling where only a signal from the
driven oscillator is available.  Similar to the method 
of Rosenblum and Pikovsky for the identification of the coupling
direction \cite{rosenblum2001b}, it is not only robust to
noise, but requires the dynamics of the driven oscillator to be
perturbed by weak internal noise.  It turns out that a regime of
``hard'' phase locking has to be distinguished from a regime of
``soft'' phase locking, and two different tests have to be used.

An advantage of the tests proposed here is that they cover all
frequency ratios $\omega_1 : \omega_2 \approx n:m$ of driven
Oscillator~1 to driving Oscillator~2 with small integers $n$ and
$m$. This includes the cases with $n=1$, where the
method described in
~\cite{janson2001}
cannot be applied.  In particular for the case of $1:1$ phase
locking, where a separation of the signals of the oscillators
involved is excluded, no test seems to be known so far.

For simplicity, the theory is first described for this important case 
only. We specify the conditions for a successful detection of phase locking,
and apply it to simulated data. Finally, a demonstration of the method for vortex flow meters is reported.

In most experiments not the phase itself but some oscillatory
signal $x(t)$ is measured. One possibility to extract the phase
$\phi(t)$ from
such a signal is to first calculate the Hilbert transform $x_{H}$ of the
signal, 
\begin{equation}
x_{H}(t)=\frac{1}{\pi}P.V.\int_{-\infty}^{\infty}\frac{x(t')}{t-t'}\,dt'
\label{def:hilbert}
\end{equation}
where $P.V.$ denotes the Cauchy principal value. Then, the \emph{analytical signal}
\begin{equation}
\zeta(t)=x(t)+ix_{H}(t)=A(t)\exp(i\phi(t))
\label{def:anasig}
\end{equation}
gives a consistent way to define the phase $\phi(t)$
\cite{pikovsky2001,rossberg2004}. By extracting the phase $\phi(t)$, 
the higher dimensional dynamics  of an oscillating system is projected
to one dimension.

A minimal model \cite{pikovsky2001} that generally describes the phase dynamics of 
two unidirectionally coupled, self-sustained oscillators is given by:
\begin{subequations}
  \label{dphi}
  \begin{align}
    \label{dphi1}
    \dot\phi_1-\omega_1&=-K \sin(\Delta \phi)+\xi_1
    \\
    \label{dphi2}
    \dot\phi_2-\omega_2&=\xi_2\quad,
  \end{align}
where $\xi_1$ and $\xi_2$ represent Gaussian noise with 
\begin{align}
  \label{noise-strength}
  \left<
    \xi_k(t)\xi_l(t')
  \right>&=2 D_k \delta_{k,l}\delta(t-t'),\quad k,l=1,2 \quad.
\end{align}
\end{subequations}
Unidirectional coupling of two self-sustained oscillators in the
presence of noise is controlled essentially by the four characteristic
time scales of the phase dynamics, which correspond to the four rate constants entering the minimal model. These are the detuning $\Delta \omega=\omega_1-\omega_2$ between the oscillators, the diffusion coefficients $D_1$ and $D_2$ of the phases $\phi_1$ and $\phi_2$ of the two
oscillators in the uncoupled limit, and the relaxation rate $K$
of the relative phase $\Delta \phi(t):=\phi_1(t)-\phi_2(t)$, which
measures the strength of the coupling.

Phase locking occurs when $|\Delta \omega|\lesssim K$.  The
coupling has a noticeable effect on the dynamics of 
Oscillator~1 only when $D_1\le \mathcal{O}(K)$, otherwise its
dynamics is blurred by the internal noise.

\begin{figure}[t]
  \centering
  \includegraphics[width=1.0\columnwidth,keepaspectratio,clip]{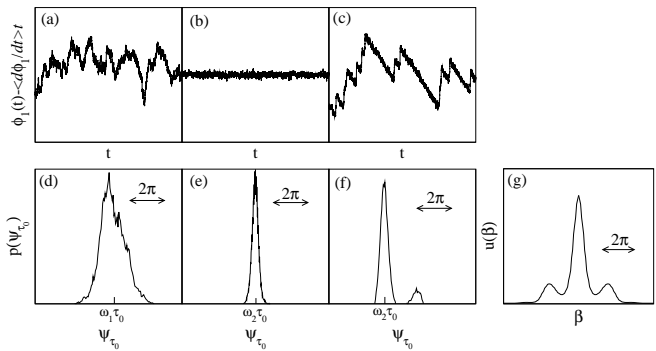}
  \caption{Typical, mean drift corrected, phase
    $\phi_{1}(t)-\left<d\phi_{1}/dt\right>t$ simulated with equation
    [\ref{dphi}] for (a) no, (b) hard and (c) soft phase locking and
    the distributions of the phase increment $\psi_{\tau_0}$ for each
    case (d)-(f). (g) the distribution $u(\beta)$, cf.~Eq.~(\ref{def:q}),
    corresponding to (f).} 
  \label{fig:dists}
\end{figure}

The minimal model for the phase dynamics, Eq.~(\ref{dphi}), can exhibit
three different types of dynamical behavior: 
\begin{itemize}
\item
If there is no phase
locking, $\phi_{1}(t)$ performs a random walk with a mean drift
$\left<d\phi_{1}/dt\right>t\sim\omega_1 t$, see
Fig.~\ref{fig:dists}a.
\item
If
the phase is locked, it can either be \emph{hard-locked}, i.e.~it
fluctuates around $\left<d\phi_{1}/dt\right>t\sim\omega_2 t$, see
Fig.~\ref{fig:dists}b,

\item
or it is \emph{soft-locked}, i.e.~the
fluctuation of $\phi_{1}(t)$ around
$\left<d\phi_{1}/dt\right>t\sim\omega_2 t$ is intercepted by rapid
increases or decreases of $2 \pi$, see Fig.~\ref{fig:dists}c. These jumps
are called \emph{phase slips}. 
\end{itemize}
These different behaviors only occur for $D_2 \ll D_1$ which is
assumed in the following. Otherwise, $\phi_{1}(t)$ always performs a
random walk.

Denoting the average number of phase slips per unit time by $r$,
the regime for soft phase locking is, expressed in the characteristic
time scales,
\begin{align}
 \label{def:soft}
 D_2\ll r\ll K&\quad,
\end{align}
whereas for the regime of hard phase locking it is 
\begin{align}
  \label{def:hard}
  r \le \mathcal{O}(D_2)&\quad.
\end{align}

Based on these properties of the minimal model, the test for phase
locking is performed in two steps: A first test 
checks for the frequent occurrence of phase slips. If the outcome
of this test is negative, a second test is performed to distinguish
the case of hard phase locking from the absence of synchronization.

In order to perform the first test, the phase increment
\begin{align}
  \label{eq:deltaphi}
  \psi_\tau(t):=\phi_1(t+\tau)-\phi_1(t)
\end{align}
over time intervals of length $\tau$ is considered.
First, an estimate of the probability distribution function
$p(\psi_{\tau_0})$ for some fixed delay $\tau_0$   is determined.
Phase slips in $\phi_1(t)$ generally lead to a multimodal
distribution $p(\psi_{\tau_0})$ with maxima separated by $2\pi$ as
displayed in Fig.~\ref{fig:dists}f, if $\tau_0$ is chosen to be longer than the
duration of a phase slip $\mathcal{O}(K^{-1})$
\cite{stratonovich1967} and shorter than the
coherence time of Oscillator~2:
\begin{align}
  \label{limits-on-tau0}
  K^{-1}\ll\tau_0\ll D_2^{-1}\quad.
\end{align}
The estimate of $p(\psi_{\tau_0})$ can be rather noisy. In order to
obtain a smooth estimator that conserves the possible multimodality of 
 $p(\psi_{\tau_0})$, the auto-covariance function
 $u(\beta)$  of $p(\psi_{\tau_0})$ is considered: 
\begin{align}
  \label{def:q}
  u(\beta):=\int_{-\infty}^{\infty} p(\psi_{\tau_0}) p(\psi_{\tau_0}+\beta)\,d\psi\quad,
\end{align}
Multimodality in $p(\psi_{\tau_0})$ translates
to multimodality in $u(\beta)$, see  Fig.~\ref{fig:dists}g, but the maxima
are now located at the positions $\beta=2\pi k$ ($k\in\mathbb{Z}$).

The degree to which $u(\beta)$ has such a multimodal structure can be
quantified by the \emph{raw phase-slip index}

\begin{equation}
  \label{def:eta0}
  \eta_{raw}:=\frac{1}{u(0)}\sum_{k=-\infty}^\infty 
  u\!\left(2k\pi\right)-u\!\left((2k-1)\pi\right)\quad.
\end{equation}
In the case of soft phase locking, determined by the relation between
the characteristic time scales of the minimal model given in
Eq.~(\ref{def:soft}), $u(\beta)$ can be approximated by a sum of
Gaussians

\begin{equation}
u(\beta)=\frac{1}{\sqrt{2\pi} \sigma} \sum_{l} a_{l}(\tau_0)
\exp\left(-\frac{(\beta-2\pi l)^{2}}{2 \sigma^2}\right) \quad, 
\end{equation}
where $a_{l}(\tau_0)$ gives the probability that in the time interval
$\tau_0$ a net number of $l$ phase slips occur, and $\sigma$ is
the width of the Gaussians. 
If the internal noise is weak, i.e.~$D_{1} \ll K$, these
Gaussians are distinct and the contribution of the minima 
$u((2k-1)\pi)$ in Eq.~(\ref{def:eta0}) can be neglected. With $\sum_{l}
a_{l}(\tau_0)=1$ this leads to an upper bound
$\eta_{max}=a_{0}(\tau_0)^{-1}$ for $\eta_{raw}$. Assuming a Poisson
distribution for the phase jumps, $a_{l}(\tau_0)$ can  be
derived analytically in dependence of the average number of phase
slips per unit time $r$ \cite{paperinprep}:

\begin{equation}
a_{l}(\tau_0)=\exp(-2r\tau_0)\,I_{l}(2r\tau_0)\quad,
\end{equation}
with $I_{l}(.)$ denoting the modified Bessel function of the first
kind \cite{abramowitz1972}.  The time delay $\tau_0$ of the phase
increment $\psi_{\tau_0}$ has to be chosen large enough that a
noticeable number of phase slips occurs. At the same time it has to be
small enough that the multimodality of $u(\beta)$ is not blurred by
the phase diffusion of the driving oscillator. Numerical simulations
show that choosing $\tau_0$ such that the Gaussian peak at the origin
contributes $\approx 75\%$ to $u(\beta)$ is a good choice. With
$a_{0}(\tau_0)=0.75$, it follows $\eta_{max}=1.33$.  Analytical
calculations \cite{paperinprep} reveal that in the case of soft
locking the value of $\var(\beta)$, with $\beta$ distributed as
$u(\beta)$, can be estimated as $8\pi^2 r\tau_0$. Thus from fixing
$a_{l}(\tau_0)$ we obtain
\begin{align}
        \label{varu}
        \var(\beta)=12.2 \quad.
\end{align}
Since the actual value of $r$ is not known, Eq. (\ref{varu}) is used
as the condition for choosing $\tau_0$.

By definition the lowest conceivable value for $\eta_{raw}$ is
obtained when $p(\psi_{\tau_{0}})$ and consequently $u(\beta)$ are single
Gaussians. With the fixed variance of $u(\beta)$
this gives a minimal value $\eta_{min}=0.014$ \cite{paperinprep} for $\eta_{raw}$ and a
\emph{normalized phase-slip index} $\eta$ can be defined as 
\begin{align}
  \label{def:eta}
  \eta:=\frac{\eta_{raw}-\eta_{min}}{\eta_{max}-\eta_{min}}\quad,
\end{align}
which takes values between 0 and 1 depending on how well
pronounced phase slips are. Large values indicate soft phase locking.
In the case of hard phase locking, no $\tau_0$ can be determined
such that Eq. (\ref{varu}) 
holds, and $\eta$ cannot be calculated.

\begin{figure}[t]
  \centering
  \includegraphics[width=0.95\columnwidth,keepaspectratio,clip]{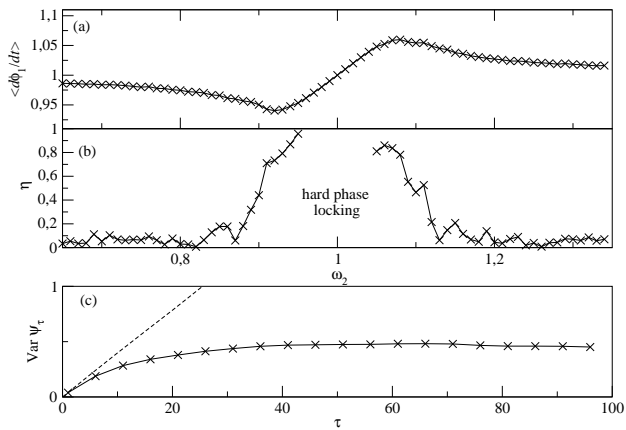}
  \caption{Simulation results.  (a) Observed frequency of driven oscillator in
  dependence on frequency of driving oscillator. 
 (b) Phase-slip index $\eta$ in dependence on driving
    frequency $\omega_2$.  (c) Variance of  $\var \psi_\tau$
 \textit{vs.} $\tau$.  The saturation of $\var \psi_\tau$ 
     indicates hard phase locking at
    $\omega_2=1.0$ (solid) compared to the uncoupled case (dashed).}
  \label{fig:OmegaPlot}
\end{figure}

If the test for soft phase locking is negative, a second test has to
be performed.  Absence of phase slips can have two reasons: Either
oscillator~1 is not synchronized to oscillator~2 at all or hard phase
locking is present. As a means to decide between the two alternatives,
the dependence of $\var \psi_\tau$ on $\tau$ is considered.  For hard
phase locking, the phase $ \phi_1(t)$ is described by a stochastic
process trapped in a local minimum of the $2\pi$-periodic interaction
potential of the two oscillators \cite{risken1989}.  In the simplest
case this is an Ornstein-Uhlenbeck process \cite{risken1989} with
$\cov [ \phi_1(t),\phi_1(t+\tau)]=(D/K) \exp(-\tau\,K)$, where
$D=D_1+D_2\approx D_1$. It follows that two regions in the dependence
of $\var\psi_\tau$ on $\tau$ can be distinguished.  For small $\tau$
there is an approximately linear increase with $\var \psi_\tau\approx
2 D_1 \tau$.  This saturates at $\tau\approx K^{-1}$ and, for larger
$\tau$, $\var \psi_\tau= 2 D_1/K$ is constant.  Such a saturation is
the signature of hard phase locking.  If oscillator 1 is not
synchronized at all, $\phi_1(t)$ performs a random walk, and the
variance of $\psi_{\tau}$ grows linearly in $\tau$ with slope $2D_{1}$
\cite{einstein1905}.

When the test for soft phase locking is applied to the general $n:m$
case, the phase $\phi_1(t)$ obtained from the signal has to be
multiplied by $m$ prior to computing $\psi_{\tau_0}$.  
In the test for hard phase locking, no modifications are required.

The proposed procedure is exemplified by a simulation study based on
the minimal model, Eq.~(\ref{dphi}), with parameters $K=0.1$,
$D_1=0.02$, $D_2=0$, $\omega_1=1$, and $\omega_2$ varied between
$0.65$ and $1.35$. Fig.~\ref{fig:OmegaPlot}a shows the observed
frequency $\left<d\phi_1/dt\right>$ in dependence on $\omega_2$,
Fig.~\ref{fig:OmegaPlot}b the normalized phase-slip index $\eta$ in
dependence on $\omega_2$, calculated from samples of $\phi_1(t)$ of
length $N=10^6$ with 10 samples per unit time. For $\omega_2$ between
$0.96$ and $1.04$, $\eta$ cannot be calculated since the condition
given in Eq. (\ref{varu}) cannot be fulfilled.
Fig.~\ref{fig:OmegaPlot}c displays the dependence of the variance of
$\psi_{\tau}$ on $\tau$. For the hard phase locked case at
$\omega_2=1$, $\psi_{\tau}$ saturates for large values of $\tau$,
while for the uncoupled case with $K=0$, the variance increases
linearly.

\begin{figure}[t]
  \centering
  \includegraphics[width=0.5\columnwidth,keepaspectratio]{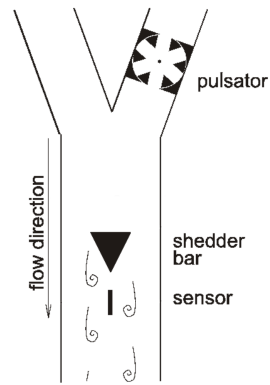}
  \caption{Experimental setup: A pulsator generates a flow pulsating
  with frequency $f_{puls}$. Downstream vortices are formed by a
  shedder bar. The frequency of the vortices $f_{vort}$ is determined via
  a piezoelectric sensor.} 
  \label{fig:expSetup}
\end{figure}

\begin{figure}[t]
  \centering
  \includegraphics[width=0.8\columnwidth,keepaspectratio]{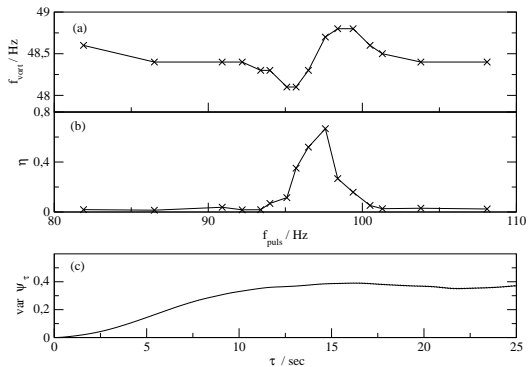}
  \caption{Experimental results.  (a) Vortex frequency  and 
    (b) phase-slip index $\eta$ in dependence on pulsation frequency
    $f_\text{puls}$.  (c) The saturation of $\var \psi_\tau$
    \textit{vs.}  $\tau$ within the region of hard phase locking at a
    larger pulsation amplitude.}
  \label{fig:XOmegaPlot}
\end{figure}

We apply the proposed procedure to detect phase locking in vortex flow meters
 that are applied to non-invasively measure flow velocity of fluids \cite{alasmi1992}.  
These devices make use of a von K{\'a}rm{\'a}n vortex street that
forms behind a shedder bar inserted orthogonal to the flow in a pipe. The
velocity of the fluid can be determined from the frequency of vortex-pair
formation $f_\text{vort}$.  In the device used here, a piezoelectric
sensor  inserted downstream behind the shedder
bar is used to detect the pressure oscillations generated from
vortices passing by.  A common problem of 
vortex flowmetering is phase locking of the vortices to pulsatile flow.

In the experiments, pulsations with frequency $f_{puls}$ were
generated by adding a periodically blocked flow to a steady flow.
The degree of flow modulation could be adjusted by the relative
contribution of the two, see Fig.~\ref{fig:expSetup}.  
In a first experiment, the peak-to-peak flow modulation 
was $\approx 10\%$ of the average flow
rate, and  the frequency of the flow
pulsation $f_\text{puls}$ was varied.  Due to the symmetry of the setup
\cite{rossberg2004}, the strongest phase locking occurs at a
frequency ratio of $f_\text{puls}:f_\text{vort}=2:1$.  In
Fig.~\ref{fig:XOmegaPlot}a, the shift in $f_\text{vort}$ due to phase
locking at $f_\text{puls}\approx 2\,f_\text{vort}$ is clearly visible.

The normalized phase-slip index $\eta$, Eq.~(\ref{def:eta}), was
computed from $65\,\mathrm{s}$ time series of the sensor signal.  As
shown in Fig.~\ref{fig:XOmegaPlot}b, the presence of phase locking is
indicated by a pronounced rise in $\eta$.  Phase locking was always
soft.
In a second experiment, pulsation was increased to $\approx 40\%$. The
condition to calculate $\eta$ were not fulfilled. 
Figure~\ref{fig:XOmegaPlot}c displays the $\var \psi_\tau$ in
dependence on $\tau$. The saturation for large values of $\tau$ indicates
 hard phase locking.

In summary, based on a minimal model for the phase dynamics of
unidirectionally coupled oscillators, we proposed a two-step procedure
that provides sensitive and specific 
indicators for soft, hard, and no phase locking. Soft phase locking is
detected by the presence of phase slips, hard phase locking by a suppression
of phase diffusion, while no phase locking is indicated by divergent
phase diffusion.  
Both methods were demonstrated on experimental data from vortex flow
meters.

The authors thank F.~Buhl and P.~Riegler for providing the flow-meter
data and the German BMBF for generous support (grant 13N7955).

\end{document}